\documentclass[twocolumn,showpacs,preprintnumbers,amsmath,amssymb]{revtex4}

\usepackage{graphicx}
\usepackage{dcolumn}
\usepackage{bm}

\begin{document}

\preprint{APS/123-QED}

\title{The Synthesis and Characterization of 1111-type Diluted Magnetic Semiconductors (La$_{1-x}$Sr$_x$)(Zn$_{1-x}$TM$_x$)AsO (TM = Mn, Fe, Co)}

\author{Jicai Lu$^{1}$, Huiyuan Man$^{1}$, Cui Ding$^{1}$, Quan Wang$^{1}$, Biqiong Yu$^{1}$, Shengli Guo$^{1}$, Hangdong Wang$^{2}$, Bin Chen$^{2}$, Wei Han$^{3}$, Changqing Jin$^{3}$, Yasutomo J. Uemura$^{4}$ and F.L.
Ning$^{1,}$}\email{ningfl@zju.edu.cn}

\affiliation{$^{1}$Department of Physics, Zhejiang University,
Hangzhou 310027, China}\affiliation{$^{2}$Department of Physics,
Hangzhou Normal University, Hangzhou 310016,
China}\affiliation{$^{3}$Beijing National Laboratory for Condensed
Matter Physics and Institute of Physics, Chinese Academy of
Sciences, Beijing 100190, China}\affiliation{$^{4}$Department of
Physics, Columbia University, New York, New York 10027, USA}

\date{\today}


\begin{abstract}
Following the synthesis of (La$_{1-x}$Ba$_x$)(Zn$_{1-x}$Mn$_x$)AsO
diluted magnetic semiconductor in Ref. 17, we investigate the doping
effect of Sr and transition metals Mn, Fe, Co into the direct gap
semiconductor LaZnAsO. Our results indicate that the single phase
ZrCuSiAs-type tetragonal crystal structure is preserved in
(La$_{1-x}$Sr$_x$)(Zn$_{1-x}$TM$_x$)AsO (TM = Mn, Fe, Co) with the
doping level up to $x$ = 0.1.  While the system remains
semiconducting, doping Sr and Fe results in a spin glass like state
below $\sim$6 K with a saturation moment of $\sim$0.02 $\mu_B$/Fe,
an order of magnitude smaller than $\sim$0.4 $\mu_B$/Mn of Sr and Mn
doped sample. The same type of magnetic state is observed neither
for (Zn,Fe) substitution without carrier doping, nor Sr and Co doped
specimens.

\end{abstract}


\maketitle

\section{\label{sec:level1}Introduction}

The successful fabrication of III-V (Ga,Mn)As ferromagnetic
semiconductors has generated great interests in the research into
diluted magnetic semiconductors (DMSs) \cite{Ohno}. Tremendous
efforts have been paid to improve the Curie temperature $T_C$, and
to understand the mechanism of the ferromagnetic ordering
\cite{Samarth,Chambers, Jungwirth}. While much progress have been
made, some inherent difficulties are encountered. For instance, the
synthesis of bulk (Ga,Mn)As specimens is difficult, because the
mismatch of Mn$^{2+}$ atoms and Ga$^{3+}$ atoms results in severely
limited chemical solubility, i.e., Mn $\leq$ 1 \% \cite{Jungwirth}.
The thin-film form specimens fabricated by the non-equilibrium
molecular-beam epitaxy (MBE), on the other hand, suffer from the
difficulty to determine precisely the amount of Mn that substitutes
ionic Ga, which donates a hole and act as a local moment, since some
Mn impurities enter the interstitial sites where they act as double
donors \cite{Jungwirth}. In diluted magnetic oxides (DMOs) such as
Co-doped ZnO and TiO$_2$, the ferromagnetism has been observed in
(Zn,Co)O and (Ti,Co)O$_2$ thin films, but not in bulk specimens,
leaving behind controversial about the origin of ferromagnetic
ordering \cite{Chambers}.

In the family of II-VI semiconductors with the form of A$^{II}$
B$^{VI}$, where A$^{II}$ = Cd$^{2+}$, Zn$^{2+}$, Hg$^{2+}$, and
B$^{VI}$ = Se$^{2-}$, Te$^{2-}$, S$^{2-}$, the substitution of
magnetic Mn atoms for A$^{II}$ can be as high as 86\% in bulk
specimens (for example, (Zn,Mn)Te \cite{Pajaczkowska,Furdyna})
because Mn$^{2+}$ is iso-valent to A$^{II}$, and the chemical
solubility is much higher. The carrier density is usually low, i.e.,
$\sim$ 10$^{17}$/cm$^3$ \cite{Wojtowicz,Morkoc}, and difficult to be
enhanced. The magnetic state in (A$^{II}$,Mn) B$^{VI}$ DMSs is
usually a spin glass with the magnetic moment size in the order of
0.01 $\mu$$_B$ per Mn atom or even smaller \cite{Furdyna,Shand}.

Recently, through the substitution of Mn for Zn and introducting
extra Li atoms in the I-II-V direct-gap semiconductor LiZnAs, Deng
et al. successfully synthesized a bulk Li(Zn$_{1-x}$Mn$_x$)As DMS
with $T_C$ as high as $\sim$ 50 K \cite{Deng1}. I-II-V Li(Zn,Mn)As
DMS has many similarities and differences to both III-V (Ga,Mn)As
and II-VI (A$^{II}$,Mn) B$^{VI}$ DMSs. The parent semiconductor
LiZnAs has a direct band gap of $\sim$ 1.6 eV
\cite{Bacewicz,Kuriyama1, Kuriyama2,Wei}, which is comparable to
that of GaAs (1.42 eV) but much smaller than 2.80 eV of ZnSe
\cite{Furdyna}; LiZnAs has a cubic structure, analogous to the
Zinc-blende structure of GaAs. Furthermore, Mn$^{2+}$ atoms are
supposed to replace the iso-valent Zn$^{2+}$, which could overcome
the chemical solubility limit encountered in III-V GaAs and InAs
systems. Another advantage is that in Li(Zn,Mn)As, the concentration
of Li can be used to precisely control the doping characteristics,
allowing possibility of either p-type or n-type doping. This
contrasts with (Ga,Mn)As which only permits p-type carriers, and
(A$^{II}$,Mn) B$^{VI}$ DMSs whose carrier density is difficult to
control.

More recently, two types of bulk DMSs with a two dimensional crystal
structure have been successfully fabricated. Firstly, Zhao et al
doped Mn and K into a ``122"-type semiconductor BaZn$_2$As$_2$, and
fabricated a DMS with $T_C$ $\sim$ 180 K \cite{Zhao}, which is
already comparable to the record $T_C$ of 190 K of (Ga,Mn)As
\cite{Wang}. Secondly, Ding et al doped Ba and Mn atoms into a
``1111"-type semiconductor LaZnAsO with a direct gap of $\sim$1.5
eV, and synthesized another DMS with $T_C$ $\sim$ 40 K \cite{Ding}.
The saturation moment size in (La,Sr)(Zn,Mn)AsO DMS is as large as 1
$\mu$$_B$/Mn, indicating the strong ferromagnetic correlation. In
A$^{II}$ B$^{VI}$, the substitution of Fe$^{2+}$ and Co$^{2+}$ for
A$^{II}$ has also induced the same type of spin glass state as that
of Mn$^{2+}$ \cite{Furdyna,Mycielski}. It will be interesting to
investigate whether Fe$^{2+}$ and Co$^{2+}$ substitution for
Zn$^{2+}$ in the two dimensional LaZnAsO are similar or different to
the case of Mn$^{2+}$ doping.

In this paper, we report the synthesis and characterization of
(La$_{1-x}$Sr$_x$)(Zn$_{1-x}$TM$_x$)AsO (TM = Mn, Fe, Co) as a
comprehensive work of previously reported (La,Ba)(Zn,Mn)AsO DMS
\cite{Ding}. We find that doping 10 \% Sr and Mn induces
ferromagnetic order with a much sharper and clearer transition at
$T_C$ $\sim$ 30 K, and a parallelogram-shaped hysteresis loop with a
coercive field of 0.158 Tesla, which is much smaller than the value
of 1.14 Tesla of the same amount of Ba and Mn doped case. Without
carriers doping, doping Fe atoms alone results in a paramagnetic
ground state. Only mediated by the carrier arising from (La,Sr)
substitution, can Fe spins magnetically order at $\sim$ 6 K. This
situation is similar to the case of Sr and Mn doping. On the other
hand, the ordered moment size in
(La$_{1-x}$Sr$_x$)(Zn$_{1-x}$Fe$_x$)AsO is as small as 0.02
$\mu_B$/Fe, resulting in the ground state more like a spin glass
than a long range ferromagnetic ordering. Furthermore, we do not
observe magnetic ordering when doping Sr and Co into LaZnAsO.

\section{\label{sec:level1}Experimental Methods}

We synthesized the polycrystalline specimens
(La$_{1-x}$Sr$_x$)(Zn$_{1-x}$TM$_x$)AsO (TM = Mn, Fe, Co) by solid
state reaction method. The intermediate products of LaAs and TMAs
(TM = Mn, Fe, Co) were prior synthesized by solid state reaction
method with La, TM, As at 900 $^{\circ}$C in evacuated silica tube
for 10 hours. The mixture of LaAs, TMAs, ZnO and SrO with nominal
concentration were heated up to 1150 $^{\circ}$C slowly and held for
40 hours in evacuated silica tube before cooling down at a rate 10
$^{\circ}$C/h to room temperature. The polycrystals were
characterized by X-ray diffraction carried out at room temperature
and dc magnetization by Quantum Design SQUID. The electrical
resistance was measured on sintered pellets with typical four-probe
method.

\section{\label{sec:level1} Results and Discussions}
\subsection{Results of (La$_{1-x}$Sr$_x$)(Zn$_{1-x}$Fe$_x$)AsO}
We show the crystal structure and X-ray diffraction pattern of
(La$_{1-x}$Sr$_x$)(Zn$_{1-x}$Fe$_x$)AsO specimens in Fig. 1. The
peaks of LaZnAsO can be well indexed by a ZrCuSiAs-type tetragonal
crystal structure with a $P4/nmm$ space group. We reproduce the
Rietveld analysis of the parent compound LaZnAsO from
ref.\cite{Ding} and show it in Fig. 1(d). Sr and Fe doping does not
change the crystal structure and the single phase is preserved with
the concentration up to $x$ = 0.10. The nonmagnetic impurity phases
of SrO$_2$ and Zn$_3$(AsO$_3$)$_2$ emerge at the level of $x$ = 0.12
and 0.15, as marked by the stars in Fig. 1(c). The a-axis lattice
parameters remain constant and the c-axis lattice parameters
monotonically increase with Sr and Fe doping up to $x$ = 0.15,
indicating the successful solid solution of (La,Sr) and (Zn,Fe).

\begin{figure}[b] \centering \vspace*{-0.1cm}
\centering
\includegraphics[width=3in]{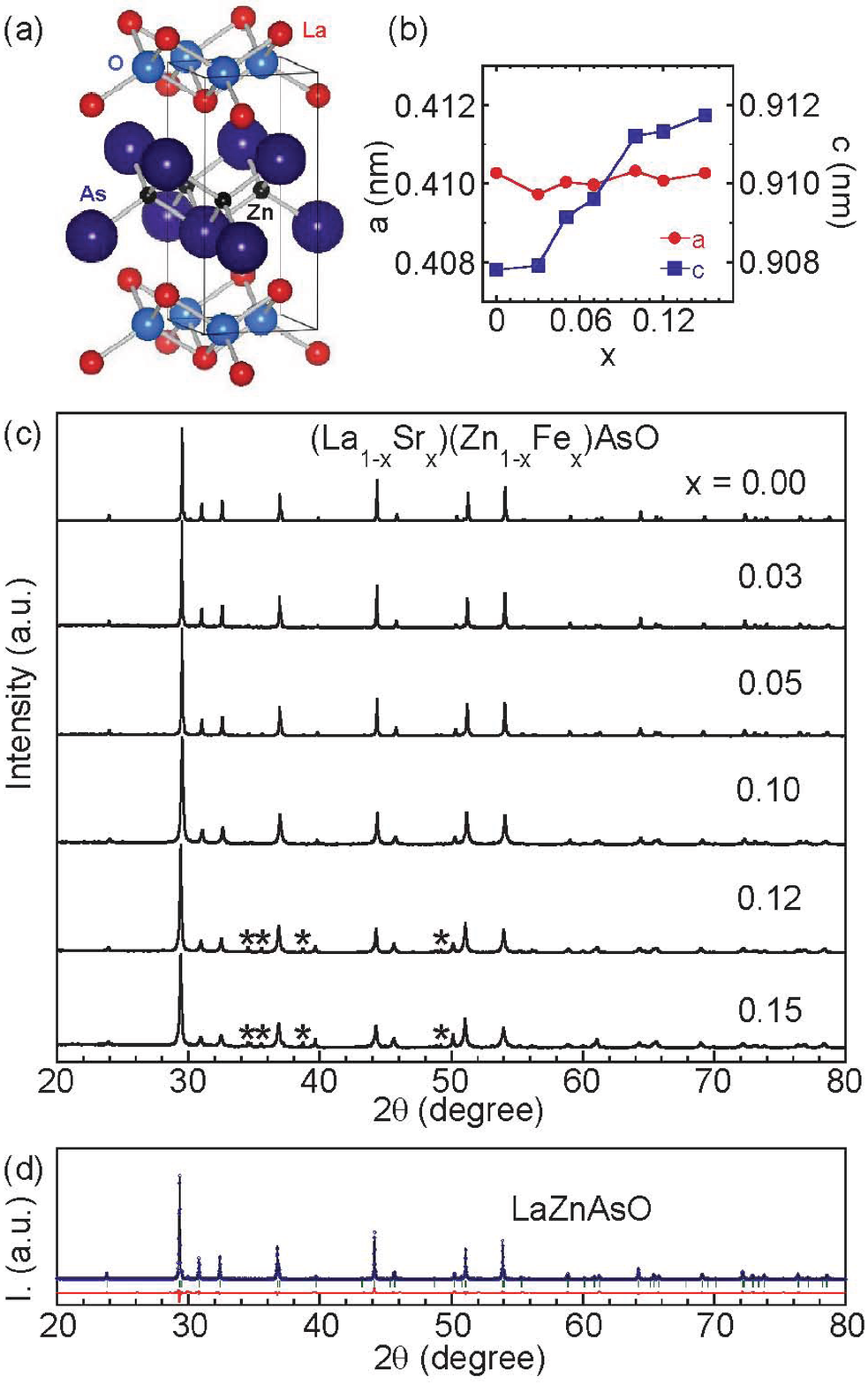}\vspace*{+0.1cm}
\caption{\label{Fig1:epsart} (Color online) (a) Crystal Structures
of LaZnAsO (P4/nmm)  (b) Lattice constants for the a-axis (red
filled circle) and c-axis (blue filled square) of
(La$_{1-x}$Sr$_x$)(Zn$_{1-x}$Fe$_x$)AsO for various doping levels
$x$. (c) X-ray diffraction pattern of
(La$_{1-x}$Sr$_x$)(Zn$_{1-x}$Fe$_x$)AsO with $x$ = 0.00, 0.03, 0.05,
0.10, 0.12, 0.15. Traces of impurity SrO$_2$ and Zn$_3$(AsO$_3$)$_2$
 are marked by stars ($\ast$) for $x$ $\geq$ 0.10. (d) X-ray diffraction
pattern of LaZnAsO with Rietveld analyses reproduced from Fig. 1(d)
of Ref. \cite{Ding}.}
\end{figure}

In Fig. 2, we show the electrical resistivity measurement for
(La$_{1-x}$Sr$_x$)(Zn$_{1-x}$Fe$_x$)AsO. The resistivity increases
with decreasing temperature down to 4 K for all doping levels,
indicating the semiconducting behavior. The temperature dependence
of resistivity are similar to the case of Sr and Mn doped
(La$_{1-x}$Sr$_x$)(Cu$_{0.925}$Mn$_{0.075}$)SO specimens
\cite{Yang}, but contrast to the case of Mn doped CaNiGe and CaNiGeH
where a Kondo like behavior is observed \cite{Liu}. The magnitude of
resistivity decrease with higher doping levels, indicating more
carriers are introduced. We do not observe the insulator to metal
transition with the doping level up to 15 \%, as oppose to the
observation in BaZn$_2$As$_2$ \cite{Zhao} and BaMn$_2$As$_2$
\cite{Bao}, where 5 \% potassium doping will induce a
metal-insulator transition. In our case, the introduction of Mn
atoms is to help the localization of the hole carriers. We also
point out that the temperature dependence of resistivity for $x$ =
0.10 sample is apparently different to other doping levels, and the
reasons are unknown. Efforts to generate single crystals are
underway to clarify this issue.

\begin{figure}[!htpb] \centering \vspace*{+1cm}
\centering
\includegraphics[width=3.1in]{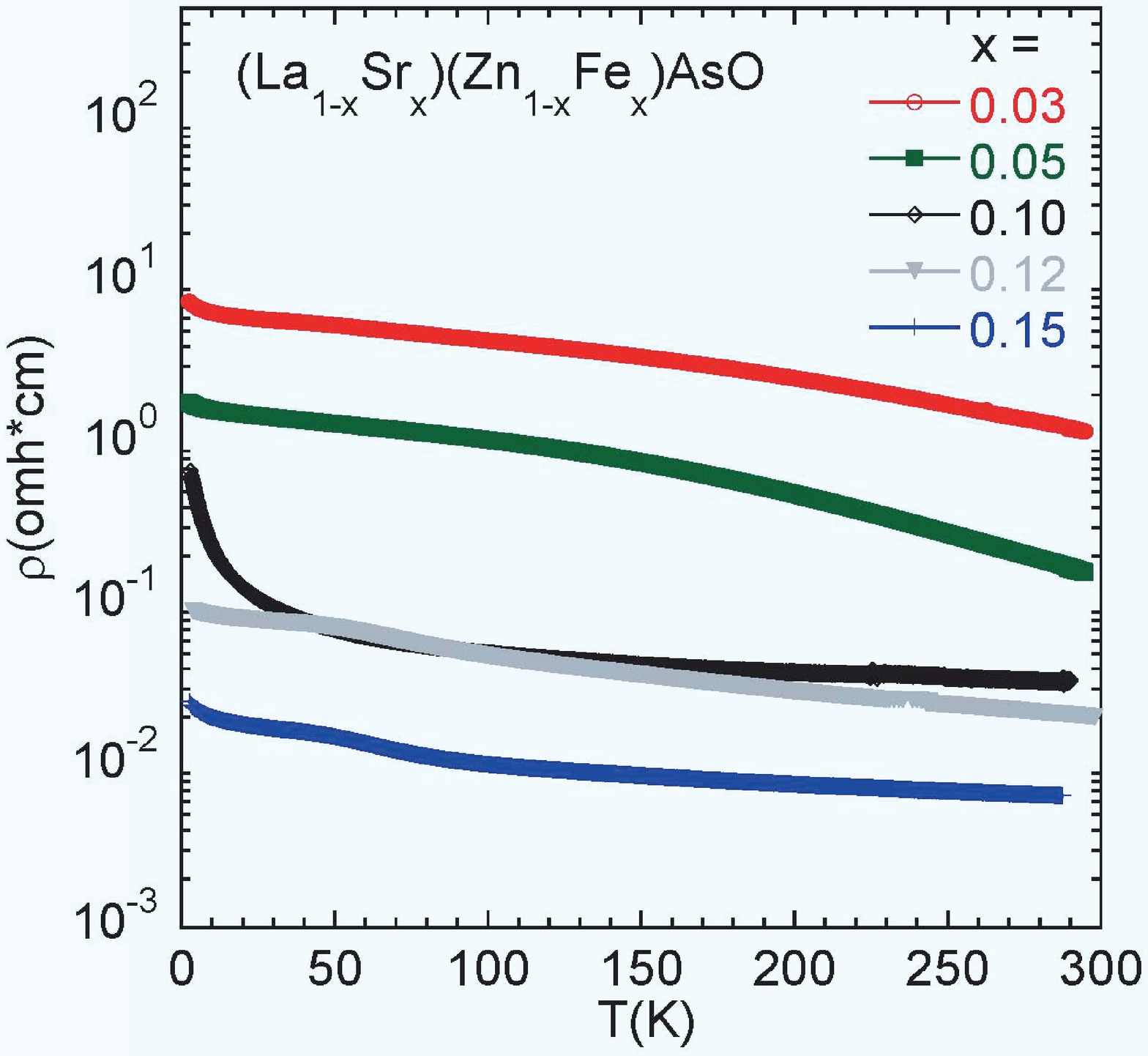}\vspace*{+0.5cm}
\caption{\label{Fig2:epsart} (Color online) Electrical resistivity
for (La$_{1-x}$Sr$_x$)(Zn$_{1-x}$Mn$_x$)AsO with $x$ = 0.00, 0.03,
0.05, 0.10, 0.12, 0.15.}
\end{figure}

In Fig. 3(a), we show the dc-magnetization of
La(Zn$_{0.9}$Fe$_{0.1}$)AsO specimen under zero-field-cooled (ZFC)
and field-cooled (FC) condition for $B_{ext}$ = 1000 Oe. We do not
observe any anomaly or splitting of ZFC and FC curve in the measured
temperature range. The curve can be well fitted by a Curie-Weiss
function with C = 0.016 $\mu$$_B$K/Fe and $\theta$ = 1.35 K,
suggesting a paramagnetic ground state. This indicates that doping
10 $\%$ Fe atoms alone does not induce any type of magnetic
ordering. The situation is similar to the case of doping Mn atoms
alone into LaZnAsO where a paramagnetic ground state is observed as
well. We reproduce the dc-magnetization curve of
La(Zn$_{0.9}$Mn$_{0.1}$)AsO in Fig. 4(a) for convenience.

\begin{figure}[!htpb] \centering \vspace*{+0.5cm}
\centering
\includegraphics[width=3.4in]{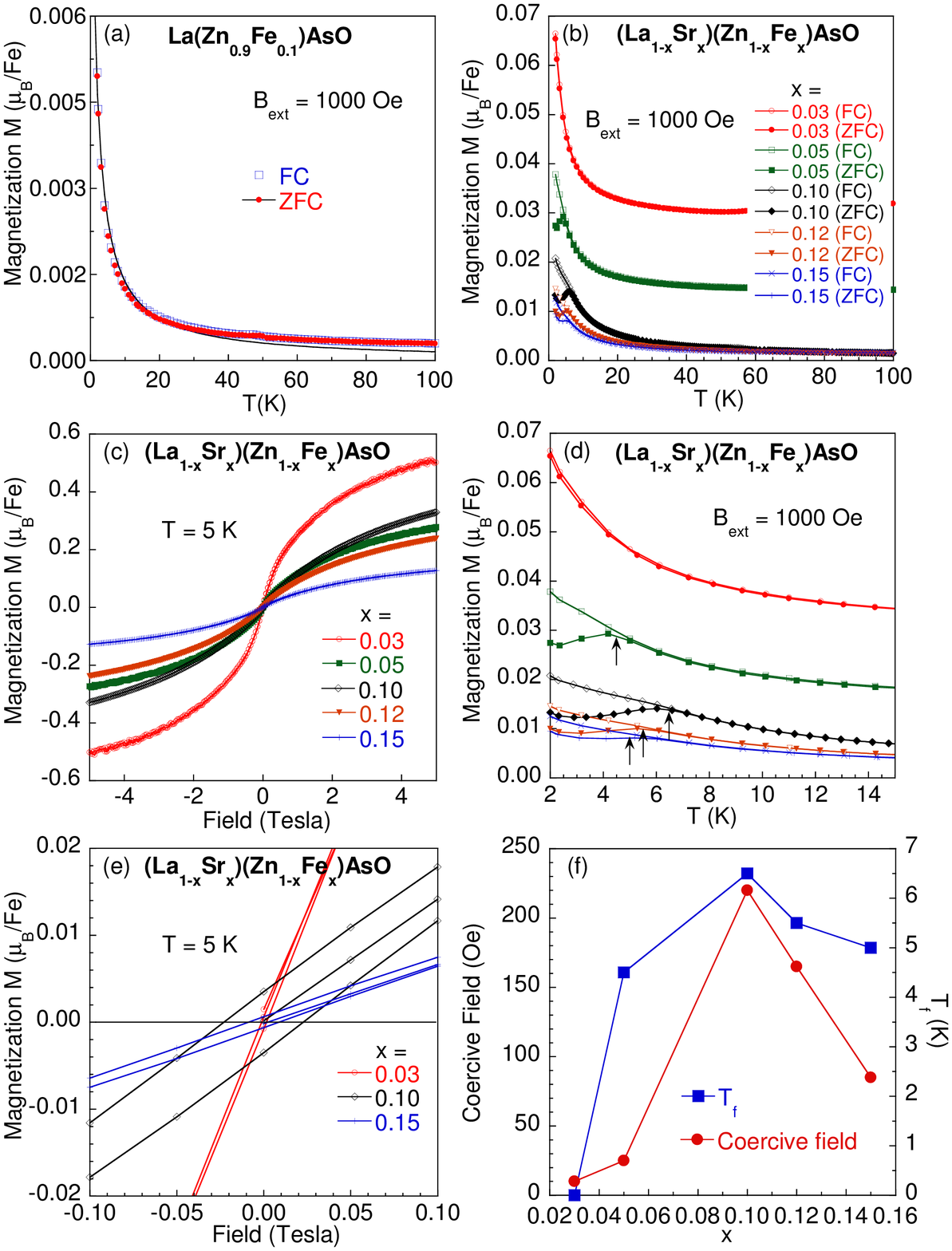}\vspace*{+0.5cm} \\
\caption{\label{Fig3:epsart} (Color online) (a) Temperature
dependence of magnetization M for LaZn$_{0.9}$Fe$_{0.1}$AsO, without
charge doping, and the solid line represents the Curie-Weiss law M =
C/(T-$\theta$) with C = 0.0156 $\mu_B$K/Fe and $\theta$ = 1.35 K
that are determined from free parameter fitting. (b) Magnetization M
under the condition of zero field cooling (ZFC) and field cooling
(FC) in the external field of 1000 Oe for
(La$_{1-x}$Sr$_x$)(Zn$_{1-x}$Fe$_x$)AsO ($x$ = 0.03, 0.05, 0.10,
0.12, 0.15). (c) The isothermal magnetization M measured at 5 K for
(La$_{1-x}$Sr$_x$)(Zn$_{1-x}$Fe$_x$)AsO; note that the hysteresis
loop is not closed even at 5 Tesla. (d) M in the temperature range
of 2 K and 15 K; arrows mark the position of $T_f$. (e) Hysteresis
loops for small field regions, which demonstrate small coercive
fields of $\sim$ 10 - 220 Oe. (f) The concentration dependence of
coercive field and $T_f$.}
\end{figure}

In Fig. 3(b), we show the dc-magnetization of
(La$_{1-x}$Sr$_x$)(Zn$_{1-x}$Fe$_x$)AsO specimens with $x$ = 0.03,
0.05, 0.10, 0.12, 0.15 under ZFC and FC condition for $B_{ext}$ =
1000 Oe. Once Sr atoms substitute for La and the carriers are
introduced into the compound, the ground state becomes different.
The magnitude of saturation moment at 2 K is about an order of
larger than that of La(Zn$_{0.9}$Fe$_{0.1}$)AsO, i.e., $\sim$ 0.07
$\mu_B$/Fe for $x$ = 0.03. For $x$ = 0.03, ZFC and FC curves are
almost superposed to each other. Starting from $x$ = 0.05, ZFC and
FC display a clear splitting at $T$ $\sim$ 4.5 K. We show the low
temperature data in the range of 2 K and 15 K in Fig. 3(d) to show
the bifurcation of ZFC and FC curves clearly. We define the
temperature where ZFC and FC curves split as $T_f$. $T_f$ increase
to 6.5 K for $x$ = 0.10 and then decrease to 5 K for $x$ = 0.15.
Contrast to the case of (La,Ba)(Zn,Mn)AsO, we do not observe a
strong enhancement of M before the bifurcation of ZFC and FC curves
that corresponds to the appearance of strong ferromagnetic
correlation .

We also measured the iso-thermal magnetization for each doping level
and show the results in Fig. 3(c). S-shape hysteresis loops are
observed for all doping levels. The magnetization is not saturated
at external field as high as 5 Tesla. We show the hysteresis loops
in small field region in Fig. 3(e), and the loops with well defined
coercive fields of 10 - 220 Oe are clearly seen. The
history-dependence behavior and the small moment size ($\sim$ 0.02
$\mu_B$/Fe) in (La$_{1-x}$Sr$_x$)(Zn$_{1-x}$Fe$_x$)AsO are similar
to the cases of typical dilute alloy spin glasses
\cite{Tholence,Monod,Prejean}, implying that a spin glass like state
has developed. In general, the spin freezing temperature $T_f$ of a
typical spin glass has DC field dependence, and frequency dependence
in the AC susceptibility measurement. In the case of
(La$_{1-x}$Sr$_x$)(Zn$_{1-x}$Fe$_x$)AsO, the cusp shape of the ZFC
magnetization is not clear, and the measurement of AC frequency
and/or DC field dependence of $T_f$ are difficult. We plot the
concentration dependence of both coercive field and $T_f$ in Fig.
3(f). Both the coercive field and $T_f$ show roughly the same
concentration dependence. They increase with the doping level up to
$x$ = 0.10, and then decrease with more Sr and Fe doping. The
decrease of coercive field and $T_f$ are likely due to the
competition of antiferromagnetic exchange interaction with
increasing Fe dopants. The same type of antiferromagnetic ordering
is observed in LaFeAsO, the parent compound for 1111-type of iron
chalcogen high temperature superconductors, at $T_N$ = 137 K
\cite{de la Cruz}.

\subsection{Results of (La$_{0.9}$Sr$_{0.1}$)(Zn$_{0.9}$Mn$_{0.1}$)AsO}

To compare with the results of
(La$_{0.9}$Sr$_{0.1}$)(Zn$_{0.9}$Fe$_{0.1}$)AsO, and refrain from
the possible influences of any other parameters, we also synthesize
the compound (La$_{0.9}$Sr$_{0.1}$)(Zn$_{0.9}$Mn$_{0.1}$)AsO. We
show the magnetization curve of
(La$_{0.9}$Sr$_{0.1}$)(Zn$_{0.9}$Mn$_{0.1}$)AsO under ZFC and FC
condition for 50 Oe in Fig. 4(b). Some features are clearly seen:
(1) M suddenly increase at Curie temperature $T_C$ $\sim$ 30 K, and
saturate to a value of 0.36 $\mu_B$/Mn at the base temperature of 2
K; (2) The ZFC and FC M-T curves split below $T_f$ $\sim$ 18 K, as
shown in Fig. 4(b); (3) Parallelogram-shaped hysteresis loop with a
well-defined coercive field of 0.158 Tesla exists, as shown in Fig.
4(c); (4) The hysteresis loop close and saturate to $\sim$
1.4$\mu$$_B$/Mn at the external field of $\sim$ 2 Tesla. The overall
features are similar to the case of (La,Ba)(Zn,Mn)AsO DMS except
that the ferromagnetic transition is much sharper and the coercive
field is much smaller in the current case. A possible reason is that
the ionic radius of Sr$^{2+}$ (0.113 nm) is much closer to that of
La$^{3+}$ (0.106 nm), and its chemical solubility is better than the
case of Ba$^{2+}$ (0.135 nm) since Sr and Ba impurities scatter
electrons strongly. The nature of $T_f$ has been determined from the
zero-field and longitudinal field -$\mu$SR measurement of
(La,Ba)(Zn,Mn)AsO. A peak of muon spin relaxation rate 1/T$_1$ is
observed at $T_f$ $\sim$ 15 K - 20 K. This indicates that the
dynamic slowing down of spin fluctuations approach to a ``static
freezing" of either individual spins or domain wall motion, as shown
in Fig. 3(f) of ref. \cite{Ding}. The ``static freezing" temperature
$T_f$ determined from the $\mu$SR data agrees well with that of the
magnetic susceptibility data measured by SQUID.

\begin{figure}[!htpb] \centering \vspace*{-1.5cm}
\centering
\includegraphics[width=3.3in]{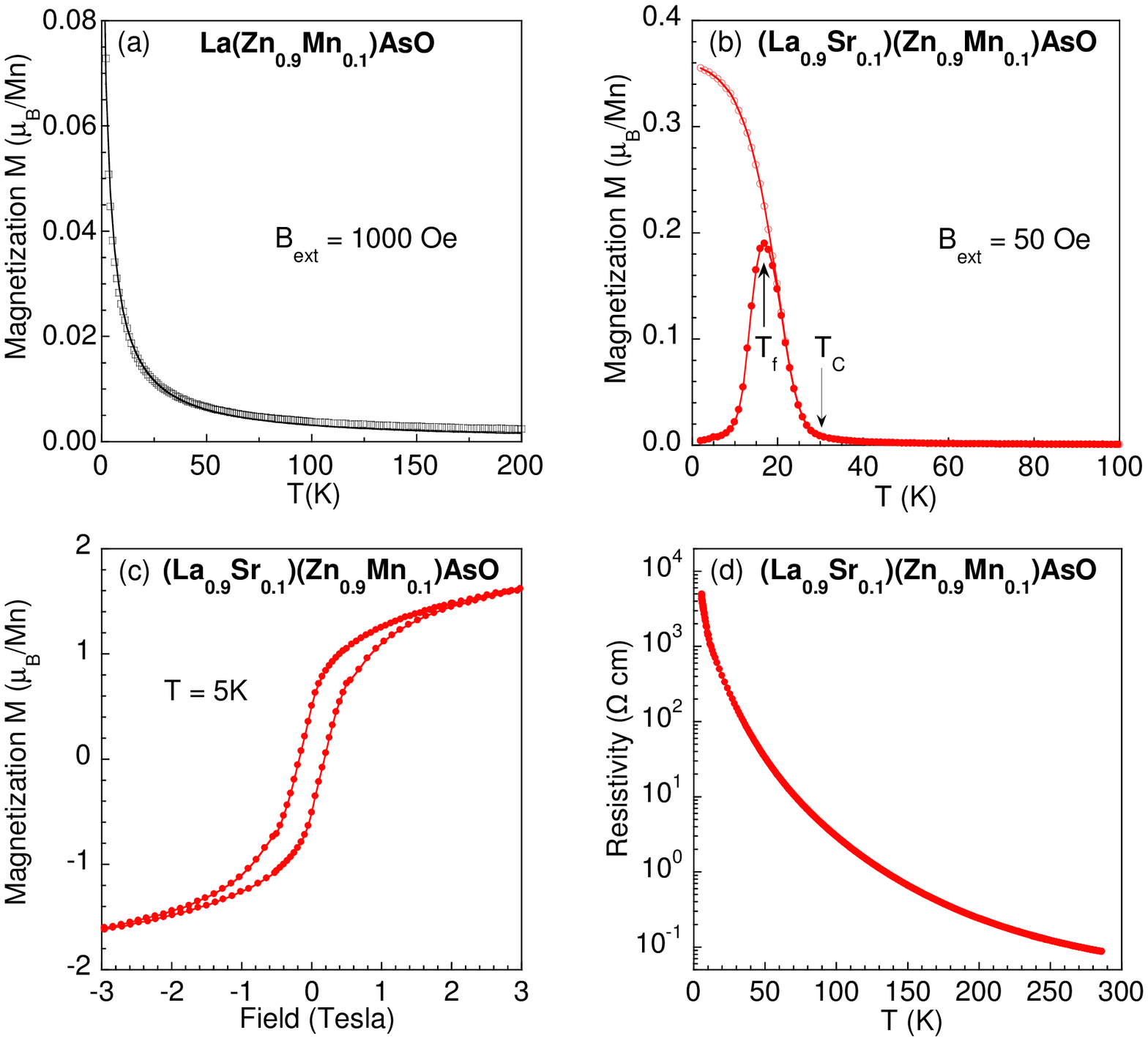}\vspace*{+2.5cm}
\caption{\label{Fig4:epsart} (Color online) (a) Temperature
dependence of magnetization M for LaZn$_{0.9}$Mn$_{0.1}$AsO, without
charge doping, and the solid line represents the Curie-Weiss law M =
C/(T-$\theta$) with C = 0.321 $\mu_B$K/Mn and $\theta$ = 2.71 K that
are determined from free parameter fitting (Reproduced from Fig.
2(a) of Ref. \cite{Ding}). (b) Magnetization M under the condition
of zero field cooling (ZFC) and field cooling (FC) in the external
field of 50 Oe for (La$_{0.9}$Sr$_{0.1}$)(Zn$_{0.9}$Mn$_{0.1}$)AsO.
(c) The isothermal magnetization M measured at 5 K for
(La$_{0.9}$Sr$_{0.1}$)(Zn$_{0.9}$Mn$_{0.1}$)AsO; the coercive field
is 1580 Oe. (d) Electrical resistivity for
(La$_{0.9}$Sr$_{0.1}$)(Zn$_{0.9}$Mn$_{0.1}$)AsO.}
\end{figure}

We have also conducted neutron diffraction measurement of
(La$_{0.9}$Sr$_{0.1}$)(Zn$_{0.9}$Mn$_{0.1}$)AsO to investigate the
nature of the ferromagnetic ordering. Unfortunately the average
moment size is only $\sim$ 0.04$\mu_B$/Mn in this diluted magnetic
system, and it is difficult to decouple the magnetic and structural
Bragg peaks even at 6 K. None the less, we find that the powder
diffraction pattern agrees well with the structure found by X-ray
shown in Fig. 1. No structure phase transition is observed in the
temperature range between 6 K and 300 K \cite{Ning}.

\subsection{Results of (La$_{0.9}$Sr$_{0.1}$)(Zn$_{0.9}$Co$_{0.1}$)AsO}
Since the saturation moment size of Fe doped sample is an order of
magnitude smaller than that of Mn doped samples, it will be
interesting to see if Sr and Co doping can induce magnetic ordering
in LaZnAsO. We show the dc-magnetization curve of
(La$_{0.9}$Sr$_{0.1}$)(Zn$_{0.9}$Co$_{0.1}$)AsO specimen in Fig. 5.
In the measured temperature range, we do not observe any anomaly or
the spin glass like state as that of
(La$_{0.9}$Sr$_{0.1}$)(Zn$_{0.9}$Fe$_{0.1}$)AsO sample. The moment
size is 0.0028 $\mu_B$/Co at 2 K. We fit the data by a Curie-Weiss
law and obtain C = 0.01$\mu_B$K/Co and $\theta$ = 1.6 K, indicating
the paramagnetic ground state. We also find that M(T) $\sim$
$T^{\alpha}$ with the power $\alpha$ = -0.78. It has been
established in II-VI DMS that a random exchange interaction in all
direction will usually give a $\alpha$ between -1 $\sim$ 0
\cite{Anderson}. The power-law behavior implies that the interaction
energy between spins is of the order of $k_B$T. The Curie-Weiss
behavior of (La$_{0.9}$Sr$_{0.1}$)(Zn$_{0.9}$Co$_{0.1}$)AsO
indicates that although Co along with Sr doping does not induce a
magnetic ordering, Co atoms do introduce local moments as Mn and Fe,
but with a much smaller moment size. We should point out that the
compound LaCoAsO is a ferromagnet with $T_C$ = 66K \cite{Yanagi}.

\begin{figure}[!htpb] \centering \vspace*{-3.5cm}
\centering
\includegraphics[width=3.3in]{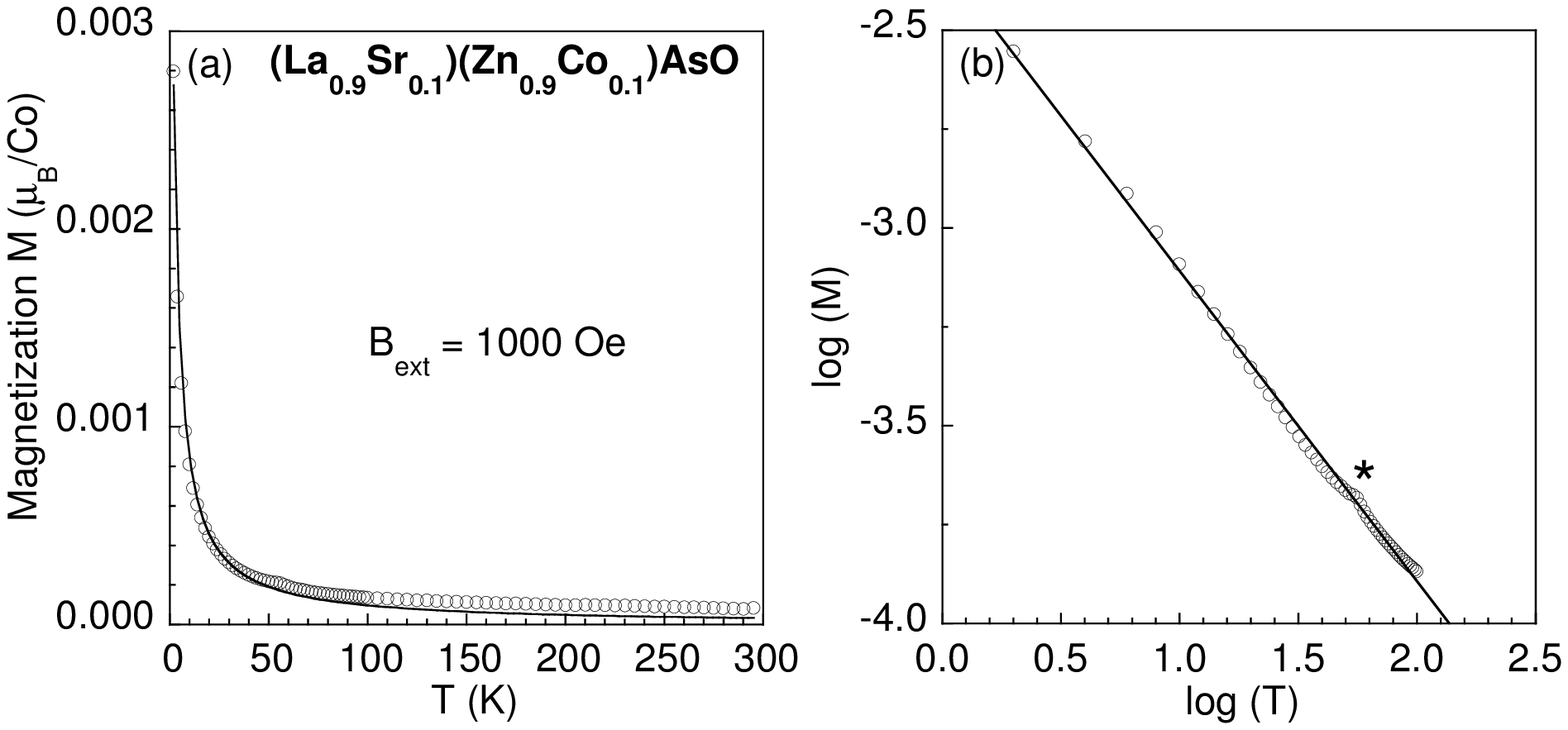}\vspace*{+4.5cm}
\caption{\label{Fig4:epsart} (a) Temperature dependence of
magnetization M for (La$_{0.9}$Sr$_{0.1}$)(Zn$_{0.9}$Co$_{0.1}$)AsO,
and the solid line represents the Curie-Weiss law M = C/(T-$\theta$)
with C = 0.01 $\mu_B$K/Co and $\theta$ = 1.62 K that are determined
from free parameter fitting. (b) Log(M) versus log(T) for
(La$_{0.9}$Sr$_{0.1}$)(Zn$_{0.9}$Co$_{0.1}$)AsO in the temperature
range of 2 K and 100 K; solid line is a linear fit to the data with
the slope of -0.78; the star marks the signal arising from the
incident oxygen during the measurement.}
\end{figure}

\section{\label{sec:level1}Summary and Discussions}

To summarize, we report the successful synthesis of polycrystalline
semiconductors (La$_{1-x}$Sr$_x$)(Zn$_{1-x}$TM$_x$)AsO (TM = Mn, Fe,
Co). The saturation moment decrease from $\sim$ 0.36 $\mu_B$/Mn to
$\sim$ 0.02 $\mu_B$/Fe, to $\sim$ 0.003 $\mu_B$/Co for the 10\%
doping level, along with the ground state from a ferromagnetic
ordered state to a spin-glass like state, and to a paramagnetic
state. The concentration of dopant and the random spacial
distribution of TM atoms are expected to be the same in each system,
but the strength of the magnetic exchange interaction decreases two
orders from Mn to Fe to Co.

In passing, ferromagnetism has been found in Fe doped SnO$_2$ ($T_C$
$\sim$ 610 K) \cite{Coey} and Co doped SnO$_2$ ($T_C$ $\sim$ 650 K)
thin films \cite{Ogale}. M\"{o}ssbauer spectra show that the iron is
all high-spin state Fe$^{3+}$ with a net ferromagnetic moment 1.8
$\mu_B$/Fe in Sn$_{0.95}$Fe$_{0.05}$O$_2$. In II-VI DMS, both iron
and cobalt are supposed to be in the valence of 2+ \cite{Furdyna,
Villeret}. In the case of (La,Sr)(Zn,Fe)AsO, supposing that Fe is in
the valence of 3+ (or $\geq$ 2+), we would expect an electron doping
once Fe$^{3+}$ substitute for Zn$^{2+}$ in
La(Zn$_{0.9}$Fe$_{0.1}$)AsO, but no ferromagnetic ordering is
observed in La(Zn$_{0.9}$Fe$_{0.1}$)AsO, as shown in Fig. 3(a). We
fit the magnetization curve above $T_f$ and find that the effective
moment is 2.37 $\mu_B$/Fe for 10\% Sr and Fe doped sample. However,
we can not estimate either the valence of Fe ions or the spin state
due to the magnetic frustration and/or the antiferromagnetic
exchange interaction in (La,Sr)(Zn,Fe)AsO. A naive expectation is
that the valence state of iron and cobalt is more like the case of
II-VI DMSs. It is also worthwhile to point out that in the ending
product LaFeAsO \cite{Hosono}, Fe$^{2+}$ is expected and its six
$d$-electrons are all important to the multi-band electronic
structures \cite{Stewart}.

We would like to emphasize that decoupling carriers and spins is the
advantages of the Sr and TM (TM = Mn, Fe, Co) doped LaZnAsO
semiconductor. Both Sr and TM dopants are chemically stable and the
concentration can be precisely controlled during the synthesis
process. The substitution of Sr for La introduces hole carriers, and
the substitution of TM for Zn introduces local moments,
respectively. Both the concentration of spins and carriers can be
finely tuned to investigate its individual influences on the
magnetic ordering. Furthermore, the
(La$_{1-x}$Sr$_x$)(Zn$_{1-x}$TM$_x$)AsO (TM = Mn, Fe, Co) DMSs
possess a common crystal structure and excellent lattice matching
with the high $T_c$ superconductor LaFeAsO$_{1-x}$F$_x$ (with $T_c$
$\sim$ 26K \cite{Hosono}), and the antiferromagnet LaMnAsO (with
$T_N$ $\sim$ 317K \cite{Emery}). There is a possibility to build up
junctions of semiconducting LaZnAsO, DMS (La,Sr)(Zn,Mn)AsO,
superconducting LaFeAsO$_{1-x}$F$_x$ and antiferromagnetic LaMnAsO.
Such junctions will naturally be an ideal object to investigate
various interesting fundamental physics properties and potential
applications for spintronics devices.

\section{\label{sec:level1}Acknowledgment}

The work at Zhejiang University was supported by National Basic
Research Program of China (No.2011CBA00103), NSF of China
(No.11274268), Zhejiang Provincial Natural Science Foundation of
China (LY12A04006) and Fundamental Research Funds for Central
Universities (2013QNA3016); F.L. Ning acknowledges partial support
by the US NSF PIRE (Partnership for International Research and
Education: OISE-0968226) and helpful discussions with X. Wan and C.
Cao.
\\



\end{document}